\documentclass[reprint,prl,twocolumn,superscriptaddress,nofootinbib,floatfix]{revtex4-1}
\usepackage{graphicx,bm,color,amsmath,amssymb}

\newcommand{\non}{\nonumber}
\newcommand{\ba}{\begin{eqnarray}}
\newcommand{\ea}{\end{eqnarray}}
\newcommand{\be}{\begin{equation}}
\newcommand{\ee}{\end{equation}}
\newcommand{\bi}{\begin{itemize}}
\newcommand{\ei}{\end{itemize}}

\newcommand{\p}{\partial}

\newcommand{\D}{\nabla}

\begin{document}

\title{Revealing Infinite Derivative Gravity's true potential: \\ The weak-field limit around de Sitter backgrounds}
\author{James Edholm}
\affiliation{Physics Department, Lancaster University, Lancaster, LA1 4YB, United Kingdom}
\begin{abstract}

General Relativity is known to produce singularities in the potential generated by a  point source. Our universe can be modelled as a de\ Sitter (dS) metric and we show that ghost-free Infinite Derivative Gravity (IDG) produces a non-singular potential 
around a dS background, while returning to the GR prediction at large distances. We also show that although there are
an apparently infinite number of coefficients in the theory, only a finite number actually affect the predictions.

By writing the linearised equations of motion in a simplified form, we find that at distances below the Hubble length scale, the difference between the IDG potential around a flat background and around a de Sitter background is negligible.

\end{abstract}
\maketitle

General Relativity (GR) has been extremely successful in describing gravity at large distances~\cite{Will}, 
in particular, the recent observation by LIGO \& VIRGO
of the collision of two neutron stars~\cite{TheLIGOScientific:2017qsa}. However, GR breaks down at short distances
 because it predicts singularities in both black holes and the cosmological setting~\cite{Hawking:1973uf,Raychaudhuri:1953yv}.

An obvious way to ameliorate these problems was to add higher derivative terms to the GR action, 
for example Stelle's 4th derivative theory~\cite{Stelle:1976gc}, 
or higher derivative gravity models~\cite{Modesto:2014eta,Giacchini:2016xns}
did not succeed because of the Ostrogradsky instability~\cite{Barnaby:2007ve},
which produces an unstable vacuum~\cite{VanNieuwenhuizen:1973fi,Woodard:2015zca,Ostrogradsky:1850fid}. 
However this instability is avoided for Infinite Derivative Gravity (IDG). IDG modifies the propagator so that it is exponentially suppressed at high energies, but doesn't produce ghosts 

Infinite derivative actions constructed from the d'Alembertian operator
$\Box=g^{\mu\nu} \nabla_\mu \nabla_\nu$, were first used in string theory~\cite{Tseytlin:1995uq}, and later employed in gravity~\cite{Biswas:2005qr}. They have been examined around a flat Minkowski background~\cite{Biswas:2011ar}
and constantly curved backgrounds~\cite{Biswas:2016etb}. 

So that the only pole in the graviton propagator is the benign pole found in GR, we can require that any infinite derivative function  in the denominator of the propagator is the \textit{exponential of an entire function}   which by definition contains no 
zeroes~\cite{Tomboulis,Siegel:2003vt,Biswas:2005qr,Biswas:2011ar,Biswas:2013kla,Biswas:2016etb,Buoninfante:2016iuf}. The mass scale $M$ of the theory determines the length scales below which the infinite derivative terms start to have a significant effect.
The prospect of avoiding singularities through IDG has prompted much investigation over recent 
years~\cite{Tomboulis,Siegel:2003vt,Biswas:2005qr,Biswas:2011ar,Biswas:2013kla,Biswas:2016etb,Buoninfante:2016iuf,Talaganis:2014ida,
Modesto:2011kw,Biswas:2005qr,
Biswas:2011ar,Edholm:2016hbt,Edholm:2017fmw,Conroy:2017nkc,Conroy:2014eja,Cornell:2017irh,Myung:2017rwf,Frolov,Frolov2015mfb,Frolov2015scs,
Biswas:2013cha,Calcagni:2013vra,Biswas:2010zk,Biswas:2012bp,Koshelev:2012qn,Koshelev:2013lfm,Biswas:2016etb,Biswas:2016egy,
Conroy2015wfa,Teimouri:2016ulk,Mazumdar:2017kxr,Modesto:2010uh,Frolov:2016xhq,Calcagni:2014vxa,Koshelev:2016vhi,
Feng:2017vqd,Calcagni:2010ab,Koshelev:2017bxd,Ghoshal:2017egr,Calcagni:2017sov,Biswas:2012bp,ArkaniHamed:2002fu,Koshelev:2017tvv,
Edholm:2016seu,Briscese:2012ys,Koshelev:2016xqb,Craps:2014wga,Buoninfante:2018xiw,Talaganis:2014ida,Talaganis:2015wva,
Talaganis:2017tnr,Conroy:2016sac,Conroy:2017nkc,Conroy:2017nkc,Conroy:2017uds,Conroy:2014dja}.

Previous work has shown that IDG gives a non-singular potential for a test mass in a flat background, which returns to the observed 
GR prediction at large distances~\cite{Edholm:2016hbt,Biswas:2011ar,Conroy:2017nkc}. Even though this calculation was made using the 
linearised equations of motion, if the test mass is small enough then the perturbation will still be in the linear regime. This is because the effect of IDG is to stop the potential from growing in size once we go reach the distances 
where IDG has a significant effect.  

We extend the prediction of a non-singular potential, which returns to GR at large distances, to a curved background. We do this by writing down the equations of motion in a simplified form and noting that in the areas we might wish to test IDG, such as laboratory experiments or solar system tests, 
we can use the approximation $H^2r^2\ll 1$~\cite{Jin:2006if}.

\section{Equations of motion}
We will investigate the action\footnote{More generally, there is also a Weyl tensor term 
$C_{\mu\nu\lambda\sigma}F_3(\Box) C^{\mu\nu\lambda\sigma}$ but it is possible to 
set $F_3(\Box)=0$ without loss of generality~\cite{Conroy:2016sac,Conroy:2017uds}. This action
is the most general torsion-free action which is quadratic in curvature.}
\ba \label{eq:action}
        S=\int d^4 x \frac{\sqrt{-g}}{2} \bigg[M^2_P R 
        + R F_1(\Box)R+ R_{\mu\nu} F_2(\Box)R^{\mu\nu} +\Lambda \bigg],~~~~
\ea
where $R_{\mu\nu}$ is the Ricci curvature tensor and $R$ is the Ricci scalar. 
The $F_i(\Box)$ are the infinite sums $F_i(\Box)=\sum_{n=0}^\infty f_{i_n}\frac{\Box^n}{M^{2n}}$,
where \{$f_{i_0},f_{i_1},f_{i_2},...,f_{i_n}$\} are the dimensionless coefficients of the series
 and 
$M$ is the mass scale of the theory. 
The lower bound on $M$ is 0.004~eV from laboratory experiments~\cite{Edholm:2016hbt,Kapner:2006si}, while the upper 
bound is the Planck mass $M_P$. Statistical analysis showed that the best fit
of the IDG\ prediction matched the data better than the GR prediction by around 2$\sigma$
\cite{Perivolaropoulos:2016ucs}.

It was shown in~\cite{Edholm:2016hbt,Biswas:2011ar,Conroy:2017nkc} that
this action generates a non-singular potential around a flat background 
for a static test mass. 
 
For the action \eqref{eq:action}, IDG gives a vacuum de Sitter solution with $\Lambda=3M^2_P H^2$~\cite{Conroy:2017uds,Biswas:2016etb}, where the background Ricci scalar is $\bar{R}=12H^2$. The linear equations of motion around this de Sitter background can be written 
as\footnote{It should be noted that this definition of $a$, $c$ and $f$ allows the minimum condition for perturbations around an (A)dS background to avoid 
Penrose singularities in \cite{Edholm:2017fmw} to be extended to include a non-zero $F_2(\Box)$.}  
\ba \label{eq:dsfieldeqns}
        \frac{1}{M^2_P} T^\mu_\nu &=& a(\bar{\Box}) r^\mu_\nu -\frac{1}{2}\delta^\mu_\nu c(\bar{\Box})r -\frac{1}{2}\D^\mu \p_\nu f(\bar{\Box})r, 
\ea   
where the functions $a(\Box)$, $c(\Box)$, $f(\Box)$, given in \eqref{acffordsfull} are combinations of $F_1(\Box)$ and $F_2(\Box)$ from the action, $r^\mu_\nu$
 and $r$ are the perturbed Ricci tensor and Ricci scalar respectively and $\bar{\Box}$ is the background d'Alembertian.
\vspace{-4mm}
\section{degrees of freedom}
\vspace{-4mm}
If $a(\Box)=c(\Box)$ then around a flat background, the propagator is given by  
\ba
        \Pi_{\text{IDG}} = \frac{\Pi_{\text{GR}}}{a(-k^2)},
\ea        
where $\Box \to -k^2$ in momentum space. We can choose $a(-k^2)\hspace{0.4mm}$=$\hspace{0.4mm}e^{\gamma(k^2/M^2)}$ where $\gamma(k^2/M^2)$ is an 
entire function. This choice means that the 
propagator has no extra poles compared to GR and also reduces to GR in the limit $M\to\infty$. 
An entire function can be written as a polynomial, i.e. 
$\gamma(k^2/M^2)=\sum_{n=1}^\infty c_n \frac{k^{2n}}{M^{2n}}$. Therefore the effect of IDG on the potential, which depends on $e^{-\gamma(k^2/M^2)}$ can be written as 
\ba
        e^{-\gamma(k^2/M^2)}= e^{-c_1 \frac{k^2}{M^2}} \cdot e^{-c_2 \frac{k^4}{M^4}}
        \cdot e^{-c_3 \frac{k^6}{M^6}}.
\ea         
This would a priori seem to have 
an infinite number of degrees of freedom as there is no restriction on the coefficients
$c_n$. It was already shown that as long as $\gamma(k^2/M^2)>0$ for $k\to \infty$ 
that IDG would produce a non-singular solution~\cite{Biswas:2011ar}. However, we can reduce the freedom in the model futher by noting that for large $n$ and positive $c_n$, 
$\exp\left[-c_n \left(\frac{k^2}{M^2}\right)^n\right]$ 
forms a rectangle function:
\ba \label{eq:rect}
        e^{-c_n \left(\frac{k^2}{M^2}\right)^n}\approx \left\{\begin{array}{lr}
        1, & \text{for } \frac{\sqrt[n]{c_n}k^2}{ M^2}< 1\\
        0, & \text{for } \frac{\sqrt[n]{c_n}k^2}{M^2}>1
        \end{array}\right\} .
\ea   

\begin{figure}[!htbp]\hspace{-4mm}\includegraphics[width=90mm]{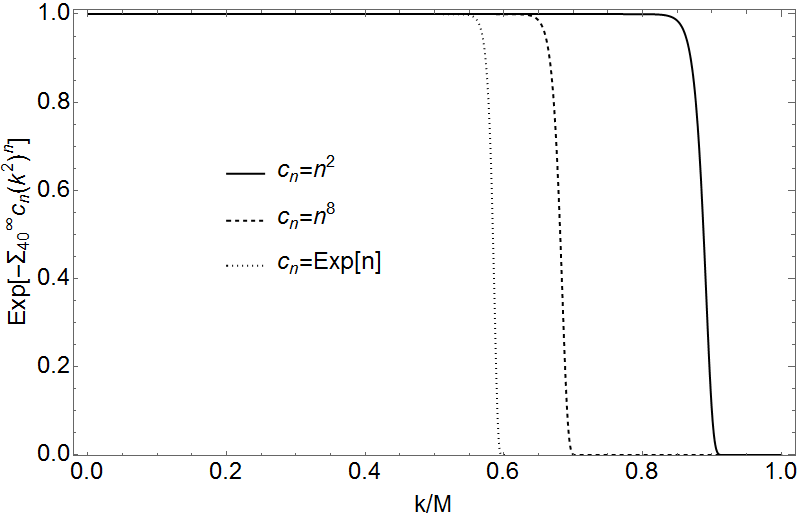}
\caption{We plot $\exp\left[-\sum_{n=40}^{\infty}c_n\left(\frac{k^2}{M^2}\right)^n\right]$ for various $c_n$ to show that even if $c_n$ 
increases very quickly, we can model the higher terms in the polynomial $\gamma$ as the rectangle
function Rect($Ck^2/M^2$) without knowing the exact values of $c_n$. Here $C$ is simply another constant to be found.}
\label{rectfct}
\end{figure}
We can combine all of the terms in the exponential which involve large values of $n$ into a rectangle function 
and thus we do not need to know the value of $c_n$ for large $n$ as long as it is positive. 
As an example, we plot $\exp\left[-\sum_{n=40}^{\infty}c_n\left(\frac{k^2}{M^2}\right)^n\right]$
for various values of $c_n$ in Fig.~\ref{rectfct}  which turns out
to still be well approximated by the rectangle function Rect($Ck^2/M^2$) even if $c_n$ 
increases exponentially. In other words, 
 the potential is
well described by the lowest order terms
and a single constant representing the higher order terms.~\footnote{We need in the region
of $\mathcal{O}(10)$ of the lower order terms depending on how 
accurate one needs to be.} 

\section{Perturbing the metric}  Following the example of~\cite{Jin:2006if}, we write the perturbation to the background de Sitter metric as 
\ba
        ds^2&=&-\left(1+2\Phi(r)-H^2 r^2\right)dt^2 \non\\
        &&+\left(1-2\Psi(r)\right)\left[\left(1+H^2r^2\right)dr^2 +r^2d\Omega^2\right],
\ea 
where $\Phi(r)$ and $\Psi(r)$ are perturbations we want to find. As we want to look at perturbations at relatively short distances such as when we are looking at laboratory 
or solar system tests of gravity, we can use the approximation $H^2r^2\ll 1$ to find
the  Ricci curvature tensors and Ricci scalar up to linear order
in $\Phi(r)$ and $\Psi(r)$
\ba \label{eq:generalperturbedmetricriccitt} 
        R_{tt}\approx \Delta\Phi(r) - 3H^2 \left(1+2 \Psi(r)-\Phi(r)\right), 
\ea
\ba  \label{eq:generalperturbedmetricriccirr}      
        R_{rr}&\approx& \frac{2 \left(r \Psi  ''(r)+\Psi  '(r)\right)}{r }-\Phi''(r)\non\\
        &&+H^2 \bigg[3-r\Phi
'(r) -2\Phi(r)+4\Psi (r)\bigg],
\ea        
\ba 
        R\approx4\Delta\Psi(r)-2\Delta\Phi (r)     
        +12 H^2  \left(1+3\Psi(r)-\Phi  (r)\right).~~~
\ea
We note that in the limit $H^2\hspace{0.3mm}r^2\ll1$, for any scalar $X(r)$ which is a function only of the radial coordinate~\cite{Jin:2006if},
\ba
        \bar{\Box}X(r)\approx \Delta\ X(r),
\ea
where $\Delta$\hspace{1mm}$\equiv$\hspace{1mm}$\eta^{ij} \D_i \p_j$\hspace{1mm}=\hspace{1mm}$\partial_r^2 \hspace{-1mm}+\hspace{-1mm}\frac{2}{r}\partial_r$ is the spatial d'Alembertian operator around a flat background.
This gives us a tremendous simplification to our equations of motion, as we will see later.

\section{Solving the equations of motion}
\begin{figure}[!htbp]\hspace{-4mm}\includegraphics[width=89mm]{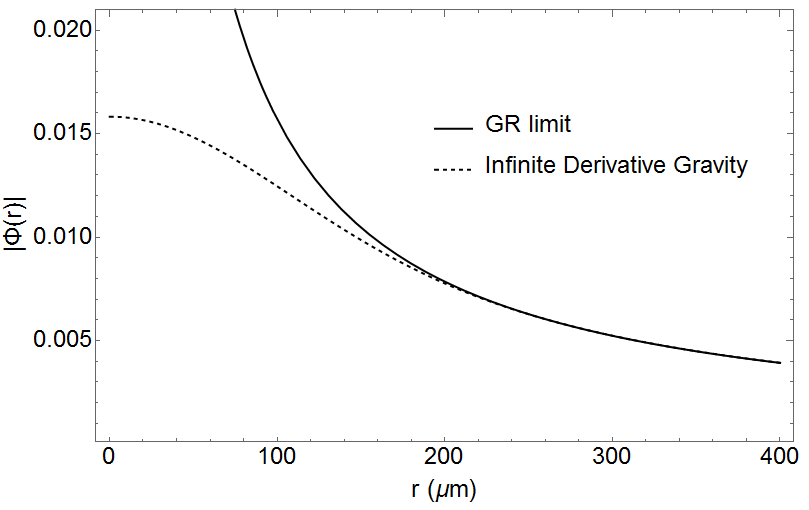}
\caption{We plot the perturbation to the metric $\Phi(r)$ versus distance $r$ using \eqref{eq:finalintegration}, giving exactly the same result as for flat backgrounds.
We have used the Hubble constant $H=7.25\times$ $10^{-27}$\hspace{0.8mm}m$^{-1}$,
set the Planck mass $M_P$ and the mass of the source equal to one 
and used the IDG mass scale $M=1.79\times10^{4}$~m$^{-1}$. 
We have chosen the coefficients $f_n$ such that around a flat background
$a$\hspace{0.8mm}=\hspace{0.8mm}$c$\hspace{0.8mm}=\hspace{0.8mm}$e^{-\Box/M^2}$, as in~\cite{Edholm:2016hbt}.  
As for flat backgrounds,  we 
have the familiar prediction $\Phi(r)\propto -\frac{\text{Erf}(r)}{r}$ \cite{Biswas:2011ar}.}
\label{Fig1}
\end{figure}

We investigate a perturbation caused by the addition of a static point mass with density $\rho=m~\delta^3(\bf{x})$ where $m$ is the mass of the source and  negligible pressure
$P\approx 0$.
Using the equations of motion and $T^\mu_\mu=-\rho=T^0_0$, we find $\Psi$ in terms of $\Phi$ in \eqref{eq:dsphiintermsofpsi}
and therefore find $\Phi$ in terms of our density $\rho$ in \eqref{Phiintermsofdensity}.

Using a similar method to \cite{Biswas:2011ar} we can go into momentum space
to find
\ba \label{eqphiintermsofintegrand}
        \Phi(r) &=&\frac{m}{4\pi^2M^2_P\hspace{0.4mm}r}\int^\infty_{-\infty}dk~I(k).
\ea   
where $I(k)$, a complicated function of $H$, $a(\Box)$, $c(\Box)$ and $f(\Box)$ is given in \eqref{eq:finalintegration}. 

\begin{figure}[!htbp]\hspace{-4mm}\includegraphics[width=89mm]{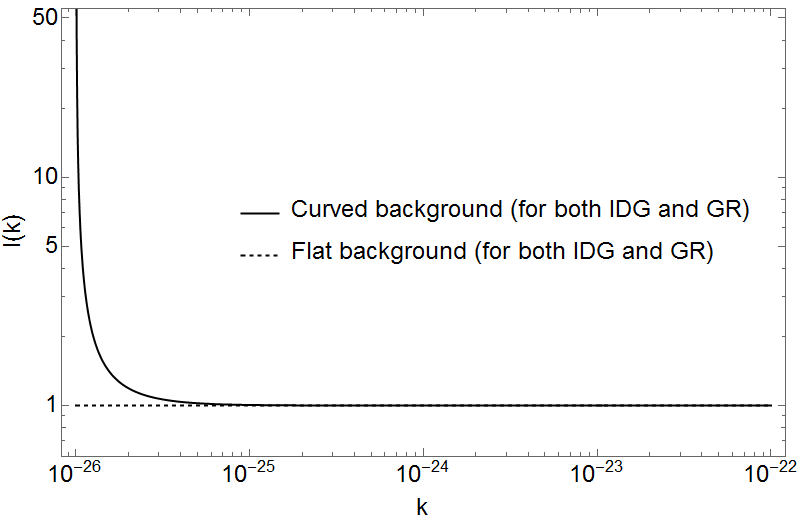}
\caption{We plot the difference from a flat background for the integrand
$I(k)$ given in \eqref{eq:finalintegration} for $H=7.25\times$ $10^{-27}$~m$^{-1}$ and $r=1$~m.
The graph is the same whether we take GR or IDG with $M=1.79\times10^{4}$~m$^{-1}$. 
Note that for $k\gg H$, $I(k)$ is similar to if we had a flat background.}
\label{smallk}
\end{figure} 

\begin{figure}[!htbp]\hspace{-4mm}\includegraphics[width=89mm]{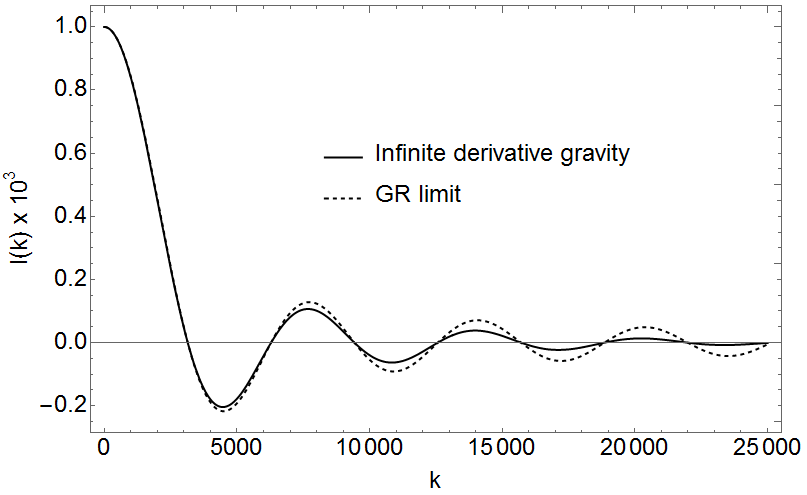}
\caption{We plot the the integrand $I(k)$ given in \eqref{eq:finalintegration}
for $H=7.25\times$ $10^{-27}$~m$^{-1}$, $M=1.79\times10^{4}$~m$^{-1}$, $r=1$m to show
that if $k<10^3$, the difference from GR is negligible.
The chosen value of $M$ is the lower bound found by experiment~\cite{Edholm:2016hbt,Kapner:2006si}
and coefficients $f_n$ such that around a flat background
$a=c=e^{-\Box/M^2}$ as in~\cite{Edholm:2016hbt}. 
For higher $M$, IDG matches GR up to even higher energy scales. The point at IDG stops being a good 
approximation for
the GR prediction of $I(k)$ is unaffected by the choice of $r$.}
\label{largek}
\end{figure}

For the value of $H$ at the current time $H$\hspace{0.8mm}=\hspace{0.8mm}$H_0=7.25\times 10^{-27}$\hspace{.4mm}m$^{-1}$, the result is very similar to the result around a flat background, 
and we plot this in Fig.~\ref{Fig1}. 
Perhaps this is to be expected - the differences from GR kick in only once we reach $k\approx M$, and our mass scale is at a significantly higher energy than the Hubble parameter, i.e. 
for large $k$, the background curvature has a negligible effect - for $H\ll k$ then the integrand is the same as for a flat background. We illustrate this by 
plotting in Fig.~\ref{smallk} and Fig.~\ref{largek} the value of the integrand for small and large $k$, choosing our value of $M$ to be the 
lower bound given by experiments~\cite{Edholm:2016hbt}. 

Note that increasing $r$ decreases the value of $k$ in the region where it
 provides the biggest contribution to the integral.\footnote{Due to the $\sin(k r)$ term, 
 the part of the integrand where $k\ll r^{-1}$ can be neglected.} Therefore for distances above
 the length scale where IDG contributes, around $10^{-5}$\hspace{.7mm}m, we can use the GR prediction, and for distances 
 at this length scale or below, we can use the IDG prediction around a flat background. 
 Therefore we still obtain a non-singular potential even when we look at a curved background. At distances larger than the mass scale of IDG, the system returns to the standard de Sitter-Schwarzschild metric.

\section{Discussion}
It was already shown that IDG gives a non-singular potential for a test mass added to a flat background,
which returns to the GR prediction at large distances.
Intuitively, one might have predicted this by noting that IDG has a significant effect
 at much larger energy scales than the 
curvature of our de Sitter universe. Here we have shown explicitly that in the region where
the background curvature must be taken into account IDG has no effect, and where IDG has an effect the 
background curvature has no effect. 

If we want to examine a system where $H$ is much larger than it is today, so that the background curvature can no longer be discarded, then it is possible to solve \eqref{eqphiintermsofintegrand} numerically, as long as $H^2r^2\ll 1$ still holds.

Could our result be extended to more general backgrounds? 
It is extremely difficult to use the equations of IDG where we encounter derivatives in more than
one coordinate, 
but it is reasonable to guess that if the background curvature is at an energy much smaller
than the energy scale where IDG becomes important, then we will see a similar effect.
\vspace{-0.2cm}
\section{Conclusion}
We have shown that the non-singular IDG potential that is found around a flat background 
also extends to a  curved de Sitter background. 
It still reverts to the GR limit at large distances. We also show that only a finite number of the degrees of 
freedom in the theory actually contribute to the potential. 

IDG is a promising model because it is ghost-free while allowing the possibility of avoiding singularities
as well as providing a framework for inflation. This result shows that we can generate a 
non-singular potential even when we are looking at non-flat backgrounds, and hints 
that we will see similar effects when looking at more general backgrounds,
based on examining the energy scales involved.


\section{Acknowledgements}
We would like to thank David Burton, Aindri\'u Conroy and David Sloan for their invaluable advice
and suggestions in preparing this paper.


\section{Appendix}
\appendix
\section{Commutation relations}
We want to write the linear equations of motion around a constantly curved background 
which is maximally symmetric, i.e.
$\bar{R}^\mu{}_{\nu\rho\sigma}=\frac{\bar{R}}{12}\left(\delta^\mu_\rho  g_{\nu\sigma}
 -\delta^\mu_\sigma g_{\nu\rho}\right)$ and 
 $\bar{R}_{\mu\nu}=\frac{\bar{R}}{4}g_{\mu\nu}$.
 
We use the relations from the appendix of \cite{Biswas:2016egy} and also derive that for a scalar $S$,
\ba 
        && \nabla^\mu F(\Box) \nabla_\nu S = F(\Box-5H^2) \nabla^\mu \nabla_\nu S \nonumber\\
        &&+ \frac{1}{4} \left(F(\Box+3H^2) -F(\Box-5H^2)\right) \delta^\mu_\nu \Box S,
\ea
and that for a symmetric tensor $t_{\mu\nu}$
\ba \label{eq:commutatorofdwithfboxforranktwotensor}
         \nabla_\mu F(\Box) t^{\mu\nu} 
        =F\left(\Box +5H^2\right) \nabla_\mu t^{\mu\nu} - 2 H^2 X(\Box)\nabla_\nu
        t^\mu_\mu,~~~~~~
\ea      
where we have defined
\ba
        X(\Box) \equiv \sum^\infty_{n=1}
        f_n \sum_{m=0}^{n-1}
        \left[\left(\Box+5H^2\right)^m \left(\Box-3H^2\right)^{n-1-m}\right].~~~~~~~
\ea
\vspace{1mm}
\section{Full equations of motion}
Using the commutation relations above, we can write the equations of motion as \eqref{eq:dsfieldeqns} with 
\ba \label{acffordsfull}
        &&a(\Box)\equiv 1 + 24  M^{-2}_p H^2  \tilde{f}_{1_0}+(\Box-2H^2)
 M^{-2}_p F_2 (\Box),\nonumber\\
        &&c(\Box)\equiv1+   M^{-2}_p \Big\{24  H^2 \tilde{f}_{1_0}-4(\Box+3H^2)F_1
(\Box)\nonumber\\
       &&- \frac{1}{2} F_2(\Box+8H^2)\Box- \frac{1}{2} F_2(\Box)(\Box+8H^2)+4H^2
\Box F_2'(\Box)\Big\}, \nonumber\\
        &&   f(\Box)\equiv  M^{-2}_p
  \left(4 F_1
(\Box)+ 2F_2(\Box) -8H^2 X(\Box-5H^2) \right),~~~~~~
\ea        
and we can therefore find $\Psi$ in terms of $\Phi$
\ba \label{eq:dsphiintermsofpsi}
        \hspace{-2.3mm}\frac{\Psi(r)}{\Phi(r)}\hspace{-0.7mm}=\hspace{-0.7mm}\frac{a(\Delta)(  \Delta     
         + 3H^2)-\left(3c(\Delta)+\Delta f(\Delta)\right)\left(\Delta   
 +6 H^2  \right)}{a(\Delta)(4  \Delta   + 30H^2)-2\left(3c(\Delta)+\Delta
f(\Delta)\right)\left(\Delta+9 H^2  \right)}, ~~~~~~
\ea             
and therefore find $\Phi$ in terms of our density $\rho$
\ba \label{Phiintermsofdensity}
        \hspace{-4.9mm}\Phi(r)\hspace{-1mm}=\hspace{-1mm}\frac{\rho\left( a(\Delta)(
 2\Delta
  + 15H^2)-\left(3c(\Delta)+\Delta
f(\Delta)\right)\left(\Delta+9 H^2  \right)\right)}
{M^2_Pa(\Delta)\left[2a(\Delta)-4c(\Delta)-\Delta f(\Delta)\right]\left(  \Delta ^2    
         + 15H^2\Delta +63H^4\right)}.\hspace{0.4mm}~~~~~
\ea
By going into momentum space, i.e. sending $\Delta\to-k^2$, taking a Fourier transform and using $\rho=m\hspace{0.8mm}\delta^3(\textbf{x})$, we can write $\Phi(r)$ as an integral we can solve
\begin{widetext}
\ba \label{eq:finalintegration}
        \Phi(r) &=&\frac{m}{4\pi^2M^2_Pr}\int^\infty_{-\infty}dk~ k\sin(kr)\frac{a(-k^2)(-4k^2
  + 30H^2)-2\left(3c(-k^2)-k^2
f(-k^2)\right)\left(-k^2-18 H^2  \right)}{a(-k^2)\left(2a(-k^2)-4c(-k^2)+k^2 f(-k^2)\right)\left(  k^4    
         -15H^2k^2 +63H^4\right)}.
\ea     
\end{widetext}
which is written in the main text as \eqref{eqphiintermsofintegrand}.


\end{document}